\documentstyle[aps,preprint,eqsecnum]{revtex}
\begin{document}
\tighten
\draft
\title{Mutual Exclusion Statistics in Exactly Solvable Models\\
in One and Higher Dimensions at Low Temperatures}
\author{
Yasuhiro Hatsugai$^{1,\dagger}$\footnote{present address:
Department of Applied Physics,
 University of Tokyo,
 7-3-1 Hongo Bunkyo-ku, Tokyo 113, Japan}
Mahito Kohmoto$^{1}$,
Tohru Koma$^{2,\dagger\dagger}$,
and
Yong-Shi Wu$^{3}$
}
\address{
$^1$ Institute for Solid State Physics, University of Tokyo,
7-22-1 Roppongi, Minato-ku, Tokyo 106, JAPAN \\
$^2$ Department of Physics, Gakushuin University,
Mejiro, Toshima-ku, Tokyo 171, JAPAN\\
$^3$Department of Physics, University of Utah,
Salt Lake City, Utah 84112, U.S.A.
}
\maketitle
\begin{abstract}
We study statistical characterization of the many-body
states in exactly solvable models with internal degrees
of freedom.  The models under consideration include the
isotropic and anisotropic Heisenberg spin chain, the Hubbard
chain, and a model in higher dimensions which exhibits
the Mott metal-insulator transition.  It is shown that
the ground state of these systems is all described
by that of a generalized ideal gas of particles (called
exclusons) which have mutual exclusion statistics, either
between different rapidities or between different species. For
the Bethe ansatz solvable models, the low temperature properties are
well described by the excluson description if the degeneracies due to
string solutions with complex rapidities are taken into account
correctly.
{For} the Hubbard chain with strong but finite
coupling, charge-spin separation is shown
for thermodynamics at low temperatures. Moreover, we present
an exactly solvable model in arbitrary dimensions which, in
addition to giving a perspective view of spin-charge separation,
constitutes an explicit example of mutual exclusion statistics
in more than two dimensions.
\end{abstract}

\pacs{PACS numbers: \hfill\break
71.27.+a(Strongly Correlated electron systems), \hfill\break
71.30.+h(Metal-Insulator transitions), \hfill\break
05.30.-d(Quantum Statistical Mechanics)\hfill}

\vfill


\section{Introduction}
\label{Introduction}
\noindent
Elementary particles or excitations are usually classified either as
boson or as fermion. In recent years, however, it has been recognized
that particles with ``fractional statistics" intermediate between boson
and fermion can exist in two-dimensional \cite{Wilczekbook} or in
one-dimensional \cite{YangYang,Haldane} systems. In two dimensions, a
type of fractional statistics can be defined on the basis of the phase
factor, $\exp(i\theta)$ with $\theta$ allowed to be arbitrary,
associated with an exchange of identical particles. One has $\theta=0$
for bosons and $\theta=\pi$ for fermions. A particle obeying such
fractional statistics (with $\theta\not=0$ or $\pi$) is called as
``anyon" \cite{wil}. It is believed that the quasiparticles
and the quasiholes in the fractional quantum Hall liquids are
anyons \cite{Halperin}. Anyons can exist only
in two spatial dimensions due to the braid group structure
associated with them \cite{Braidgroup}.

Another aspect of quantum statistics involves state counting, or the
exclusive nature of the particles. Any number of bosons can be in a
single-particle quantum state. Therefore there is no exclusion between
bosons. On the other hand, the exclusion is perfect for fermions in a
sense that a single particle state can accommodate at most one
fermion. This aspect of quantum statistics can be generalized, as
noticed by Haldane \cite{Haldane}, who proposed a definite
generalization of Pauli principle such that one can consider particles
with non-perfect exclusion. He pointed out that a spinon in
one-dimensional long-range interacting quantum spin chain can exclude,
on average, half of other spinon in occupying a single particle state.
We shall call such a generalization as ``exclusion statistics" and a
particle obeying it as an ``excluson". In contrast to usual bosons and
fermions, the general concept of exclusons allows mutual statistics.
Namely, there may exist statistical interactions or mutual exclusion
between different species of particles. Haldane has recognized
\cite{Haldane} that quasiparticles in the fractional quantum Hall
fluids are exclusons with mutual statistics between quasi-electrons
and quasi-holes.

Thus the concepts of fractional anyon statistics and exclusion
statistics constitute generalizations of two different aspects
(exchange phase and exclusion) of usual quantum statistics. An
essential difference between the two concepts is that anyons can
exist only in two spatial dimensions, while in principle exclusons
may exist in any dimensions.

Recently, one of us\cite{Wu} introduced the concept of generalized
ideal gas of exclusons (see Sec.~\ref{excluson} for definition),
and showed that its thermodynamic properties can be
easily understood through a
statistical distribution that interpolates between bosons and
fermions. Later Bernard and Wu\cite{BernardWu} have shown that the
Bethe ansatz solvable models in one dimension can be described as an
ideal (or non-interacting) excluson gas.  This exemplifies that in
certain circumstances particle-particle interactions can be totally
absorbed by the statistical interactions.
(A possible relation to a conformal field theory was
discussed by Fukui and Kawakami \cite{Kawakami}.)
Thus the concept of exclusion statistics may become a powerful
tool in understanding certain interacting many-body problems.
For example, recently it has been shown \cite{Boson} that the
essential features of low-temperature physics of Luttinger
liquids in one dimension can be approximately described by a
system of noninteracting exclusons. This may provide a
new approach to interacting many-body systems.

The examples considered by Bernard and Wu are the repulsive
$\delta$-function boson gas\cite{delta} and the Calogero-Sutherland
model\cite{1/r2}. Both of them contain only single species. The
present paper is a follow-up to study statistical interactions or
mutual statistics in exactly solvable models with internal quantum
numbers.  We will discuss one-dimensional Bethe ansatz solvable
models such as the isotropic (XXX) and anisotropic (XXZ)
Heisenberg chain, and the Hubbard chain.
In addition, we will consider an exactly solvable model in
arbitrary dimensions proposed by two of us\cite{HK}, which has the Mott
metal-insulator transition.  A common feature in all these models is
that there exists mutual exclusion either between particles of
different rapidities or between different species.

In this paper we are going to address the following two questions.
First, many Bethe ansatz solvable models with internal degrees of
freedom allow solutions of the Bethe ansatz equations with complex
rapidities, while a naive analysis of Yang-Yang thermodynamic Bethe
ansatz in terms of exclusons deals only with excitations which
correspond to solutions with {\it real} rapidities. The question is
whether these excitations are sufficient to give correct thermodynamic
properties {\em at low temperatures} or not.  Previously in ref.
\cite{Kawakami} the same topic of thermodynamics in terms of exclusons
with multi-species is discussed without addressing this problem in the
Bethe ansatz solvable models. For the Heisenberg spin chain,
we explicitly show that at the isotropic antiferromagnetic point
(with the anisotropy parameter $\Delta=1$), the generalized ideal
excluson gas of excitations with real rapidities
does give correct low-temperature behavior if a double
degeneracy of the excitations induced by complex rapidities
is taken into account.
The second question deals with the excluson
description for the physically interesting phenomenon of charge-spin
separation. Both the Hubbard chain and the exactly solvable model in
higher dimensions exhibit this phenomenon under certain conditions.
We will show that when this happens, indeed the two models
can be described by two species (spin and charge) of exclusonic
excitations with nontrivial mutual statistics.  Particularly the
latter model provides an example of (mutual) exclusion statistics
in more than two dimensions. We note that our treatment of the
models differs from that presented in ref. \cite{Spalek}.

\section{Excluson Description}
\label{excluson}
\noindent
We consider a system with a total number $N=\sum_{j,\mu} N_j^\mu$
of particles or quasiparticles,
where $N_j^\mu$ is the number of particles of species $\mu$ with
a set of {\em good} quantum numbers, collectively denoted by $j$,
specifying the states.
Following \cite{Wu}, we assume that the total number of
states with $\{N_j^\mu \}$ is
\begin{equation}
W=\prod_{i,\mu} {\left[D_i^\mu(\{N_j^\nu\})+N_i^\mu-1\right]! \over
N_i^\mu!\left[D_i^\mu(\{N_j^\nu\})-1\right]!},
\label{Wnumber}
\end{equation}
where $D_i^\mu(\{N_j^\nu\})$ is the number of available single particle
states (counted as bosons), which by definition is given by
\begin{equation}
D_i^\mu(\{N_j^\nu\})+\sum_{j,\nu} g_{ij}^{\mu\nu}N_j^\nu=G_i^\mu,
\label{mutualEq}
\end{equation}
with statistical interactions $g_{ij}^{\mu\nu}$.
Here $G_i^\mu$ is the number of available single particle states
when there is no particle in the system.
Namely, $G_i^\mu=D_i^\mu(\{0\})$. The derivative of
(\ref{mutualEq}) is
\begin{equation}
{\partial D_i^\mu(\{N_j^\nu\}) \over \partial N_j^\nu}
=-g_{ij}^{\mu\nu},
\end{equation}
agreeing with the original definition for ``{\em statistical
interactions}" proposed by Haldane\cite{Haldane}. When the pair
$(i,\mu)$ differs from $(j,\nu)$, we call $g_{ij}^{\mu\nu}$
{\em mutual} statistics between particles labelled by $(i,\mu)$ and
those by $(j,\nu)$.

We further assume that the total energy of the system with
$\{N_j^\mu\}$ particles is always simply given by
\begin{equation}
E=\sum_{j,\mu}{N_j^\mu}\epsilon_j^\mu
\label{totalE}
\end{equation}
with constant $\epsilon_j^\mu$. Equations (\ref{Wnumber}),
(\ref{mutualEq}) and (\ref{totalE}) define the generalized
ideal gas of exclusons \cite{Wu}. It is known that (\ref{totalE})
is not satisfied for free anyons \cite{nosum}.
One of us \cite{Wu} has derived statistical distribution for
generalized ideal gas and the thermodynamics following from it.
The equilibrium statistical distribution for $\{N_{i}^{\mu}\}$
is determined by
\begin{equation}
w_i^\mu N_i^\mu + \sum_{j,\nu}g_{ij}^{\mu\nu}
N_{j}^{\nu} = G_i^\mu,
\label{eq_for_w}
\end{equation}
where $w_i^\mu$ satisfy the  equations
\begin{equation}
(1+w_i^\mu ) \prod _{j,\nu}
{\left({\frac {w_j^\nu}{1+w_j^\nu} } \right)}^{g_{ji}^{\mu\nu}}
= \exp \left[{\frac {\epsilon_i^\mu-a^{\mu}} {T}}\right],
\label{eq_det_w}
\end{equation}
where $a^\mu$ is the chemical potential for particles of
species $\mu$.
The thermodynamic potential is given by
\begin{eqnarray}
\Omega & \equiv & -T \log Z\\
& = & -T \sum_{\mu,i} G_i^\mu \log \left[
{
\frac
{G_i^\mu +N_i^\mu-\sum_{j,\nu} g_{ij}^{\mu\nu} N_j^\nu}
{G_i^\mu -\sum_{j,\nu} g_{ij}^{\mu\nu} N_j^\nu}
}\right],
\label{OmegaExdes}
\end{eqnarray}
where $Z$ is the grand partition function.

\section{Thermodynamic Bethe Ansatz}
\label{Betheansatz}

\noindent A large class of models which can be interpreted as
generalized ideal gas is the Bethe ansatz solvable models in one
dimension.  It is well known that the generalized ideal gas actually
represents a system of interacting particles.  The Bethe ansatz method
can be applied to systems with the special property that all the
scattering amplitudes for many quasiparticles can be written in terms
of two-body scattering amplitudes \cite{Sutherland}. As a result, the
eigenvalue problem of a Bethe ansatz solvable system is reduced to
solving the so-called Bethe ansatz equations \cite{Bethe}.

In general, the Bethe ansatz equations may be rewritten as
\begin{equation}
L p_\mu(\lambda_\ell^\mu)=2\pi I_\ell^\mu
+\sum_{m,\nu} \theta_{\mu\nu}(\lambda_\ell^\mu,\lambda_m^\nu),
\label{BetheEq}
\end{equation}
with $\theta_{\mu\nu}(\lambda_\ell^\mu,\lambda_m^\nu)$ being the the
two-body scattering phase shift between the quasiparticles with
rapidities $\lambda_\ell^\mu$ and $\lambda_m^\nu$.  Here $L$ is the
size of the system, $p_\mu(\lambda_\ell^\mu)$ is a function of the
rapidity $\lambda_\ell^\mu$ (or the quasi-momentum), and
$\{I_\ell^\mu\}_{\ell,\mu}$ is a set of integers or half-odd integers.
We assume that the integers $\{I_\ell^\mu\}$ satisfy
$I_{\ell+1}^\mu>I_\ell^\mu$. We restrict ourselves to the solutions of
Eq. (\ref{BetheEq}) with rapidities $\lambda_\ell^\mu$ {\em all real}.
It is known that solutions with complex rapidities do exist for many
of Bethe ansatz solvable systems. But we have not been able to deal
with the problem of state counting for such solutions. Below we will
address, in several concrete examples, the question of how important
are the contributions of solutions with complex rapidities to
low-temperature properties of the system.

Subtracting the Bethe ansatz equations (\ref{BetheEq}) for
adjacent rapidities $\lambda_\ell^\mu$ gives
\begin{equation}
L\left[p_\mu(\lambda_{\ell+1}^\mu)-p_\mu(\lambda_\ell^\mu)\right]
=2\pi(I_{\ell+1}^\mu-I_\ell^\mu)
+\sum_{m,\nu} \left[\theta_{\mu\nu}(\lambda_{\ell+1}^\mu,\lambda_m^\nu)
-\theta_{\mu\nu}(\lambda_\ell^\mu,\lambda_m^\nu)\right] .
\label{BetheeqD}
\end{equation}
Following \cite{LiebLiniger,YangYang}, we introduce a particle
density
\begin{equation}
\rho_\mu(\lambda_\ell^\mu)
={1 \over L(\lambda_{\ell+1}^\mu-\lambda_\ell^\mu)}
\end{equation}
and a hole density
\begin{equation}
\rho_\mu^h(\lambda_\ell^\mu)
={M_\ell^\mu \over L(\lambda_{\ell+1}^\mu-\lambda_\ell^\mu)},
\end{equation}
where $M_\ell^\mu\ge 0$ is the number of the holes in the branches
of the Bethe ansatz equations (\ref{BetheEq}), i.e.,
the number $M_\ell^\mu$ is given by
\begin{equation}
I_{\ell+1}^\mu-I_\ell^\mu=1+M_\ell^\mu .
\end{equation}
In terms of these densities $\rho_\mu$ and $\rho_\mu^h$, we can
rewrite the Bethe ansatz equations (\ref{BetheeqD}) as
\begin{equation}
{L \over 2\pi}p_\mu^\prime(\lambda_i^\mu)\Delta \lambda_i^\mu=
L\rho_\mu(\lambda_i^\mu)\Delta \lambda_i^\mu
+L\rho_\mu^h(\lambda_i^\mu)\Delta \lambda_i^\mu
+{1\over 2\pi}\sum_{j,\nu} \left[\theta_{\mu\nu}^\prime
(\lambda_i^\mu,\lambda_j^\nu)
\Delta \lambda_i^\mu \right]
L\rho_\nu(\lambda_j^\nu)\Delta \lambda_j^\nu ,
\label{Betheeq2}
\end{equation}
where we denote the derivative of a function $f$ by $f^\prime$.
Here $L\rho_\mu(\lambda_i^\mu)\Delta \lambda_i^\mu$ and
$L\rho_\mu^h(\lambda_i^\mu)\Delta \lambda_i^\mu$ are, respectively,
the number of the quasiparticles and the number of the quasihole
in the interval $\Delta \lambda_i^\mu$.
The total number of possible choices of states with densities
$\{\rho_\mu\}$ and $\{\rho_\mu^h\}$
is given by
\begin{equation}
W_{\rm TBA}=\prod_{i,\mu}
{\left\{L[\rho_\mu(\lambda_i^\mu)+\rho_\mu^h(\lambda_i^\mu)]\Delta
\lambda_i^\mu\right\}! \over
\left[ L\rho_\mu(\lambda_i^\mu)\Delta
\lambda_i^\mu \right]! \left[L\rho_\mu^h(\lambda_i^\mu)\Delta
\lambda_i^\mu\right]!}.
\label{numberTBA}
\end{equation}
This number $W_{\rm TBA}$ of states, obtained from the Bethe ansatz
equations (\ref{Betheeq2}), is of the exactly the same form as
Eq. (\ref{Wnumber}) for the number $W$ of states in a generalized
ideal gas of exclusons with the statistical interactions given by
Eq. (\ref{mutualEq}). In fact, we get the equivalence by setting
\begin{equation}
G_i^\mu={L \over 2\pi}p_\mu^\prime(\lambda_i^\mu)
\Delta \lambda_i^\mu,
\label{Ident1}
\end{equation}
\begin{equation}
N_i^\mu=L\rho_\mu(\lambda_i^\mu)\Delta \lambda_i^\mu \quad ,
\quad D_i^\mu(\{N_j^\nu\})=L\rho_\mu^h(\lambda_i^\mu)
\Delta \lambda_i^\mu,
\label{Ident2}
\end{equation}
and
\begin{equation}
g_{ij}^{\mu\nu}=\delta_{ij}\delta_{\mu\nu}+{1\over 2\pi}
\theta_{\mu\nu}^\prime (\lambda_i^\mu,\lambda_j^\nu)
\Delta \lambda_i^{\mu}.
\label{Ident3}
\end{equation}

In the thermodynamic limit, the Bethe ansatz equations
(\ref{Betheeq2}) become
\begin{equation}
{1 \over 2\pi}p_\mu^\prime(\lambda)=\rho_\mu^h(\lambda)
+\sum_\nu \int g^{\mu\nu}(\lambda,\lambda^\prime)
\rho_\nu(\lambda^\prime) d\lambda^\prime
\label{TBAeq}
\end{equation}
with the statistical interactions
\begin{equation}
g^{\mu\nu}(\lambda,\lambda^\prime)=\delta_{\mu\nu}
\delta(\lambda-\lambda^\prime)+{1\over 2\pi}
\theta_{\mu\nu}^\prime (\lambda,\lambda^\prime) .
\label{statint}
\end{equation}
The key point is that the Bethe ansatz equations (\ref{TBAeq})
are equivalent to the equations (\ref{mutualEq}) of the
statistical interactions in the excluson formalism.

Here we stress once more that in the above we have considered
only solutions with real rapidities. The equation (\ref{statint})
for statistical interactions refers only to such excitations.
But its validity is obviously independent of whether these
solutions of Bethe ansatz equations are complete or not.
Though right now we do not know how to do state counting for
solutions with complex rapidities, we feel it plausible that a
bit more complicated excluson picture may still apply.

\section{Thermodynamic Formalism at Finite Temperatures}
\label{TFFT}
\noindent

The thermodynamic formalism for Bethe ansatz solvable models
at finite temperatures was first developed by Yang and Yang
\cite{YangYang} for the cases without internal degrees of freedom.
This formalism has been justified, e.g.\ , for a one-dimensional
Bose gas with repulsive point interaction \cite{Dorlas}. In this
section, we briefly review a straightforward generalization of
the Yang-Yang thermodynamic formalism in general setting with
internal degrees of freedom, and show that it is actually the same
as generalized ideal gas with the identification
(\ref{Ident1})-(\ref{Ident3}).

In the thermodynamic limit, the number of particles of species
$\mu$ per volume is given by
\begin{equation}
{N_\mu \over L}=\int d\lambda \; \rho_\mu(\lambda) .
\end{equation}
{From} (\ref{numberTBA}), we have the entropy $S=\log W_{\rm TBA}$ as
\begin{equation}
{S \over L}=\sum_\mu \int d\lambda \left\{
\left[\rho_\mu(\lambda)+\rho_\mu^h(\lambda)\right]
\log \left[\rho_\mu(\lambda)+\rho_\mu^h(\lambda)\right]
-\rho_\mu(\lambda)\log \rho_\mu(\lambda)
-\rho_\mu^h(\lambda)\log \rho_\mu^h(\lambda)\right\}.
\label{entropy}
\end{equation}
We assume that the total energy per volume is expressed as
\begin{equation}
{E \over L}=\sum_{\mu} \int d\lambda \; \rho_{\mu}(\lambda)
\epsilon_{\mu}^0(\lambda)
\end{equation}
in terms of an energy density $\epsilon_\mu^0$. As a rule, this
assumption is always satisfied in Bethe ansatz solvable models.

The thermodynamic potential $\Omega$ at equilibrium
with temperature $T$ can be evaluated by minimizing
\begin{equation}
\Omega=E-\sum_\mu a_\mu N_\mu -TS
\end{equation}
with respect to the variation of the densities $\rho_\mu$
and $\rho_\mu^h$. Here $a_\mu$ is the chemical potential.

As a result, we have \cite{Wu,YangYang}
\begin{equation}
{\Omega \over L}=-{T \over 2\pi}\sum_\mu \int d\lambda
p_\mu^\prime(\lambda)\log \left[1+w_\mu^{-1}(\lambda)\right] ,
\label{Omega}
\end{equation}
where the functions $w_\mu$ are determined by the equations
\begin{equation}
\log [1+w_\mu(\lambda)]-\sum_\nu \int d\lambda^\prime
g^{\nu\mu}(\lambda^\prime,\lambda)\log
\left[1+w_\nu^{-1}(\lambda^\prime)\right]
={\epsilon_\mu^0(\lambda)-a_\mu \over T}.
\label{weq}
\end{equation}
The particle densities $\rho_\mu$ at equilibrium are determined
by the Bethe ansatz equations
\begin{equation}
{1 \over 2\pi}p_\mu^\prime(\lambda)=\rho_\mu(\lambda)w_\mu(\lambda)
+\sum_\nu \int g^{\mu\nu}(\lambda,\lambda^\prime)
\rho_\nu(\lambda^\prime) d\lambda^\prime .
\label{TBAeqFT}
\end{equation}
The hole densities are given by $\rho_\mu^h=\rho_\mu w_\mu$.

For the thermodynamics reviewed here, the following cautious
remark is in order. We have followed the Bethe ansatz
method \cite{Bethe} to reduce the energy eigenvalue problem in
the models to solving the so-called Bethe ansatz equations.
However, it has not yet been generally proved that all the
solutions to the Bethe ansatz equations provide the complete
set of energy eigenstates \cite{Koma}. In particular, it has been
known in a number of cases that there are solutions with
complex rapidities to the Bethe ansatz equations in addition to
those with real rapidities. Though exact thermodynamics must deal
with all eigenstates in a complete basis, low-temperature
properties of a system, which are main focus of interests in
many situations and in the present paper, might involve only a
set of solutions that are not necessarily complete.
It has been shown that the ground state at absolute zero
always corresponds to a solution with all
rapidities real. A sensible question is thus whether the
solutions with real rapidities are enough for accounting
for low-temperature properties of the system. A way
of investigating this problem is to apply the thermodynamics
reviewed above to some cases where the exact thermodynamics
has been studied by ways that avoid the completeness
assumption, and then compare the results to the exact ones.
One well-known example of such cases is the Heisenberg spin
chain, which we are going to study in next section.

\section{Isotropic and Anisotropic Heisenberg Spin Chains}
\noindent
In this section, we apply the formalism reviewed in the last
section to the isotropic and anisotropic quantum
spin-1/2 chains. On one hand, for the spin-1/2 isotropic
Heisenberg antiferromagnetic chain at low temperatures,
we are able to explicitly demonstrate the magnon
excitations (des~Cloizeaux-Pearson-Faddeev-Takhtajan
mode \cite{DCP,FT}) in the excluson description. On the other
hand, we will show that for the isotropic antiferromagnetic
chain, the low temperature behavior obtained from a
generalized ideal gas of the doubly degenerated or
spin-1/2 excitations with real rapidities does
agree with the known exact results
\cite{XXXC,Babujian} which are obtained in other ways without
invoking the completeness assumption of string solutions
\cite{KomaTBA}.
Here we stress that for getting the correct results,
it is necessary to take into account a double degeneracy
of the excitations induced by complex rapidities.
Strictly speaking, the solutions of the Bethe-Ansatz
equations excitations with only real rapidities
generally do not properly account for the low-temperature
behavior, so other solutions with complex rapidities
have to be included. This problem has not been addressed
in the previous treatment \cite{Kawakami} on
mutli-component systems in one dimension.

\subsection{Spin-1/2 Isotropic (XXX) Heisenberg Chain}
\label{XXX}
\noindent
The Hamiltonian of the spin-1/2 XXZ chain is given by
\begin{equation}
{\cal H} = \sum_{j=1}^L \left[S_j^x S_{j+1}^x + S_j^y S_{j+1}^y +
\Delta (S_j^z S_{j+1}^z-1/4) \right]
\label{hamHei}
\end{equation}
with the periodic boundary condition ${\bf S}_{L+1} = {\bf S}_1$,
where ${\bf S}_j$ is the spin-1/2 operator at site $j$, and
$\Delta$ is the anisotropy parameter.

First consider the isotropic $(\Delta=1)$ Heisenberg chain.
As is well known, for this model, Bethe \cite{Bethe}
first reduced the eigenvalue problem of the Hamiltonian to solving
the so-called Bethe ansatz equations
\begin{equation}
2L\tan^{-1}(2\lambda_\alpha)
=2\pi I_\alpha+2\sum_{\beta=1}^M
\tan^{-1}\left(\lambda_\alpha - \lambda_\beta \right),
\quad (\alpha=1,2,\cdots,M),
\label{BetheeqXXX}
\end{equation}
where $M$ is the number of down spins.
The energy eigenvalues are given by
\begin{equation}
E=-\sum_{\alpha=1}^M {2 \over 1+4\lambda_{\alpha}^2}.
\end{equation}

Comparing (\ref{BetheeqXXX}) with (\ref{BetheEq}), we have
\begin{equation}
p(\lambda_\alpha)=2\tan^{-1}(2\lambda_\alpha)
\end{equation}
and
\begin{equation}
\theta(\lambda_\alpha,\lambda_\beta)
= 2\tan^{-1}\left(\lambda_\alpha - \lambda_\beta \right).
\end{equation}
Further, from (\ref{statint}), we get the statistical interactions
between magnons as
\begin{equation}
g(\lambda,\lambda^\prime)=\delta(\lambda-\lambda^\prime)
+{1 \over \pi}{1 \over 1+(\lambda-\lambda^\prime)^2} .
\end{equation}
Yang and Yang proved \cite{YangYang2}
that the ground state at $T=0$ indeed
corresponds to a solution with all rapidities real, so that
it is the same as the ground state of a generalized ideal
gas with above statistical interactions.

\subsection{Spin-1/2 Isotropic Heisenberg Chain at Low Temperatures}
\noindent
By substituting the results in Section~\ref{XXX} into
(\ref{Omega}) and (\ref{weq}), we get the thermodynamic potential
$\Omega$ as
\begin{equation}
{\Omega \over L}= -{T\over 2\pi}
\int_{-\infty}^\infty d\lambda {4 \over 1+4\lambda^2}
\log \left[1+e^{-\epsilon(\lambda)/T}\right] ,
\label{OmegaXXX}
\end{equation}
where $\epsilon=\log w$ is determined by the equation
\begin{equation}
\epsilon(\lambda)=-{2 \over 1+4\lambda^2}+
{T \over \pi}\int_{-\infty}^\infty d\lambda^\prime
{1 \over 1+(\lambda-\lambda^\prime)^2}
\log\left[1+e^{-\epsilon(\lambda^\prime)/T}\right]
\label{eps}
\end{equation}
Further, from (\ref{TBAeqFT}),
we have
\begin{equation}
\rho(\lambda)\left[1+e^{\epsilon(\lambda)/T}\right]
={1 \over \pi}
 {2 \over 1+4\lambda^2}-{1 \over \pi}
\int_{-\infty}^\infty d\lambda^\prime \rho(\lambda^\prime)
{1 \over 1+(\lambda-\lambda^\prime)^2}.
\label{spindensity}
\end{equation}

First consider the zero temperature limit. Then
(\ref{eps}) and (\ref{spindensity}) become
\begin{equation}
\epsilon(\lambda)=-{2 \over 1+4\lambda^2}-
{1 \over \pi}\int_{-\infty}^\infty d\lambda^\prime
{1 \over 1+(\lambda-\lambda^\prime)^2}
\epsilon(\lambda^\prime)
\end{equation}
and
\begin{equation}
\rho(\lambda)={1 \over \pi}
 {2 \over 1+4\lambda^2}-{1 \over \pi}
\int_{-\infty}^\infty d\lambda^\prime \rho(\lambda^\prime)
{1 \over 1+(\lambda-\lambda^\prime)^2}.
\end{equation}
Since these equations are, respectively, linear with respect to
the unknown functions $\epsilon$ and $\rho$, one can easily obtain
the solutions as
\begin{equation}
\epsilon(\lambda)=-{\pi \over 2\cosh \pi \lambda}
\label{epsR}
\end{equation}
and
\begin{equation}
\rho(\lambda)={1 \over 2\cosh \pi \lambda}.
\label{0rho}
\end{equation}

By substituting (\ref{0rho}) into the right-hand side of
(\ref{spindensity}), i.e., by iteration, we get the expression
\begin{equation}
\rho(\lambda)={1 \over 2\cosh \pi \lambda}
{1 \over 1+e^{\epsilon(\lambda)/T}}
\label{rhoT}
\end{equation}
at low temperatures.

To clarify the physical meaning of (\ref{epsR}) and (\ref{rhoT}),
we introduce a new variable $k\in [0,\pi]$ by
\begin{equation}
\sinh \pi \lambda = -\cot k.
\end{equation}
In terms of $k$, we can rewrite (\ref{epsR}) and (\ref{rhoT})
as
\begin{equation}
\epsilon(\lambda)=\varepsilon_k=-{\pi \over 2}|\sin k|
\label{epsDCP}
\end{equation}
and
\begin{equation}
\rho(\lambda)d\lambda={1 \over 2\pi}
{dk \over 1+e^{\varepsilon_k/T}} .
\label{spinfermi}
\end{equation}
Equation (\ref{epsDCP}) implies that hole-like excitations
have the dressed energy $\varepsilon_k^h=\pi|\sin k|/2$,
which is nothing but the des~Cloizeaux-Pearson-Faddeev-Takhtajan
mode \cite{DCP,FT}. It is known \cite{DCP} that the momentum of
this mode, $q\in (-\pi, \pi]$, is related to the variable $k$ by
\begin{equation}
q =\Biggl\{ \begin{array}{lll}
\pi -k,  \;\;\; & {\rm for} & \;\;\; 0< q < \pi; \\
      - k,  & {\rm for} &  -\pi< q <0.
\end{array}
\end{equation}
This double degeneracy is induced by complex rapidities
\cite{Babujian}, whose net effect is to give spin-1/2 to
the excitations\cite{FT}. A certain explanation of this
degeneracy will be given in the case of the spin-1/2 XY
chain below. Equation (\ref{spinfermi}) implies that in terms of
the dressed energy (\ref{epsDCP}), these hole-like excitations
obey Fermi statistics at low temperatures. In fact,
by combining (\ref{spinfermi})
with (\ref{entropy}) and
$\rho_\mu^h=\rho_\mu w_\mu=\rho_\mu e^{\epsilon/T}$,
we have the expression of the entropy at low temperatures as
\begin{equation}
{S \over L}=-{1 \over 2\pi}\int_{-\pi}^\pi dq
\left\{ f(\varepsilon_q)\log f(\varepsilon_q)
+[1-f(\varepsilon_q)]\log [1-f(\varepsilon_q)]\right\}
\label{entropyXXX}
\end{equation}
with the Fermi distribution function
\begin{equation}
f(\varepsilon)={1 \over 1+e^{\varepsilon/T}}.
\end{equation}
This result agrees with the free fermion
descriptions of the XXX chain \cite{Yamada}.
The entropy (\ref{entropyXXX}) gives the specific heat
$C\cong 2T/3$ per volume at low temperatures.
The resulting specific heat is
just the known exact specific heat \cite{XXXC,Babujian}.
Although we have dropped (string) solutions
with complex rapidities to the Bethe ansatz equations,
we have been able to get the correct low temperature
behavior by taking into account the double degeneracy
of the magnon excitations induced by complex rapidities.
Strictly speaking, the contributions to the specific heat
from the string solutions cannot be ignored at low
temperatures. But the total contribution \cite{Babujian}
appear to be just doubled, due to the spin-1/2 character
\cite{FT} of the magnon excitations.
This implies that the low energy excitations as
quasiparticle can be indeed described by the
generalizad ideal gas of exclusons with real
rapidities only.
\subsection{Spin-1/2 Anisotropic (XXZ and XY) Chains}
\noindent
Next consider the spin-1/2 chains (\ref{hamHei})
with anisotropy $-1<\Delta<1$.
Then the Bethe ansatz equations are given by \cite{YangYang2}
\begin{equation}
Lp(\lambda_\ell)=2\pi I_\ell -
\sum_{m=1}^M \Theta(\lambda_\ell,\lambda_m)
\label{BetheeqXXZ}
\end{equation}
for the states with $M$ down spins. Here
\begin{equation}
e^{ip(\lambda)}={e^{i\eta}-e^\lambda \over e^{i\eta+\lambda}-1}
\label{XXZrapidity}
\end{equation}
and
\begin{equation}
\Theta(\lambda_\ell,\lambda_m)=-2\tan^{-1}
\left[(\cot \eta)\tanh\left({\lambda_\ell-\lambda_m\over 2}\right)
\right]
\end{equation}
with $\Delta=\cos\eta$, $(0<\eta<\pi)$.
The energy eigenvalues are given by
\begin{equation}
E=-\sum_{\ell=1}^M
\left[\Delta-\cos p(\lambda_\ell) \right].
\label{XXZenergy}
\end{equation}
Comparing (\ref{BetheeqXXZ}) with (\ref{BetheEq}),
and using (\ref{statint}), we get the
statistical interactions between magnons as
\begin{equation}
g(\lambda,\lambda^\prime)=\delta(\lambda-\lambda^\prime)
+{1 \over 2\pi}{\sin 2\eta \over \cosh(\lambda-\lambda^\prime)
-\cos 2\eta} .
\end{equation}
Again, the ground state at $T=0$ is
a solution with all rapidities real \cite{YangYang2},
so that it coincides with the ground state of a generalized
ideal gas with above statistical interactions.

In the special case of the XY model, i.e., $\Delta=0$ $(\eta=\pi/2)$,
we have
\begin{equation}
g(\lambda,\lambda^\prime)=\delta(\lambda-\lambda^\prime).
\label{XYstatistics}
\end{equation}
This implies that quasiparticles (magnons) obey Fermi statistics.
Of course, this is consistent with the well-known fact that
the spin-1/2 XY chain can be transformed into a system of free
spinless fermions in one dimension by using
the Jordan-Wigner transformation \cite{LSM}.

Note that $-(\pi-\eta)<p(\lambda)<(\pi-\eta) \longleftrightarrow
-\infty<\lambda<+\infty$ from (\ref{XXZrapidity}).
In particular, we have $-\pi/2<p(\lambda)<\pi/2$ for the XY chain.
Combining this with (\ref{XXZenergy}), (\ref{weq}) and (\ref{Omega}),
we have the thermodynamic potential
\begin{equation}
{\Omega \over L}=-{T \over 2\pi}
\int_{-\pi/2}^{\pi/2}dp \log \left\{1+\exp[\cos p/T]\right\}
\label{XYOmega}
\end{equation}
for the XY chain. However this {\em does not} coincide
with the known exact thermodynamic potential \cite{LSM}
which is given by replacing
the range of integral (\ref{XYOmega}) to $[-\pi,\pi]$.
One possible reason for this discrepancy is that
we have not taken into account solutions with complex
rapidities $\lambda_\ell$ to the Bethe ansatz equations
(\ref{BetheeqXXZ}) \cite{Bethe}.
But, as is well known, there is no complex rapidity
in the XY chain \cite{LSM}. Where does this inconsistency come from ?
The answer is the following. In the procedure to take the
XY limit $\Delta \rightarrow 0$, we missed
a double degeneracy of the magnon excitations.
Actually a more careful treatment taking into account
complex rapidities shows this double degeneracy
in the the XY limit \cite{TakahashiSuzuki}.
A similar situation occurs also in the XXX chain at low temperatures
\cite{Babujian}. Of course, a much simpler way to get
the exact result of the XY chain is to use the variable $p$
instead of the rapidity $\lambda$. Then one can easily
obtain the exact result.

\section{The Hubbard Chain}
\noindent

In this section we turn to the Hubbard chain. Our intention is
to present an exclusonic description for spin-charge separation,
which is known to occur in this model with infinite coupling
at zero temperature \cite{ogata_shiba}. We will demonstrate
spin-charge separation under the following two {\it broader}
conditions: (i) The on-site Coulomb energy $U$ is large but
finite; (ii) The temperature $T$ is finite but sufficiently
low compared to the order of the effective exchange $1/U$.

\subsection{General Considerations}
\label{Hubbard}
\noindent

The Hamiltonian of the Hubbard chain is given by
\begin{equation}
{\cal H}=\sum_{j=1}^L \sum_{\sigma=\uparrow,\downarrow}
(c_{j,\sigma}^\dagger c_{j+1,\sigma}+
c_{j+1,\sigma}^\dagger c_{j,\sigma})+U \sum_{j=1}^L
c_{j,\uparrow}^\dagger c_{j,\uparrow}c_{j,\downarrow}^\dagger
c_{j,\downarrow}
\end{equation}
with on-site Coulomb energy $U$
and periodic boundary conditions
$c_{L+1,\sigma}=c_{1,\sigma} \ (\sigma=\uparrow,\downarrow)$,
where $c_{j,\sigma}^\dagger,c_{j,\sigma}$ are, respectively,
the creation and annihilation operators for an electron with
spin $\sigma$ at the site $j$.

For the Hubbard chain,
Lieb and Wu \cite{LiebWu} obtained the Bethe ansatz equations
\begin{equation}
Lk_i^c=2 \pi I_i^c + \sum_{\beta=1}^M \theta\left({2 \over U}
\sin k_i^c-\lambda_\beta^s\right) , \quad (i=1,2,\cdots,N)
\label{BetheHub1}
\end{equation}
and
\begin{equation}
-\sum_{j=1}^N\theta\left(\lambda_\alpha^s-{2 \over U}\sin k_j^c\right)
=2\pi I_\alpha^s-\sum_{\beta=1}^M
\theta\left({\lambda_\alpha^s \over 2}-
{\lambda_\beta^s \over 2}\right), \quad (\alpha=1,2,\cdots,M)
\label{BetheHub2}
\end{equation}
with quasi-momenta $\{k_i^c\}_{i=1}^N$ and
rapidities $\{\lambda_\alpha^s\}_{\alpha=1}^M$
for $(N-M)$ electrons with up spin and $M$ electrons with down spin.
Here
\begin{equation}
\theta(x)= -2 \tan^{-1}(2x).
\label{theta}
\end{equation}
The energy eigenvalues are given by
\begin{equation}
E=-2\sum_{j=1}^N \cos k_j^c.
\end{equation}

Comparing (\ref{BetheHub1}) and (\ref{BetheHub2}) with
(\ref{BetheEq}), we get
\begin{equation}
p_c(k_i^c)=k_i^c ,
\end{equation}
\begin{equation}
p_s(\lambda_\alpha^s)=0 ,
\end{equation}
\begin{equation}
\theta_{cc}(k_i^c,k_j^c)=0 ,
\end{equation}
\begin{equation}
\theta_{cs}(k_i^c,\lambda_\beta^s)=\theta\left({2 \over U}\sin k_i^c
-\lambda_\beta^s\right),
\end{equation}
\begin{equation}
\theta_{ss}(\lambda_\alpha^s,\lambda_\beta^s)=-
\theta\left({\lambda_\alpha^s \over 2}-
{\lambda_\beta^s \over 2}\right),
\end{equation}
and
\begin{equation}
\theta_{sc}(\lambda_\alpha^s,k_j^c)=
\theta\left(\lambda_\alpha^s-{2 \over U}\sin k_j^c\right).
\end{equation}
Further, from (\ref{statint}), we have
\begin{equation}
g^{cc}(k,k^\prime)=\delta(k-k^\prime),
\end{equation}
\begin{equation}
g^{cs}(k,\lambda)=-{4 \over \pi U}{\cos k \over
1+4\left(2\sin k/U -\lambda\right)^2},
\end{equation}
\begin{equation}
g^{ss}(\lambda,\lambda^\prime)
=\delta(\lambda-\lambda^\prime)+{1\over \pi}
{1\over 1+(\lambda-\lambda^\prime)^2}
\end{equation}
and
\begin{equation}
g^{sc}(\lambda,k)=-{1 \over \pi}
{2 \over 1+4\left(\lambda-2\sin k/U\right)^2}.
\end{equation}
Since in the ground state all $k_{j}^{c}$
and $\lambda_{\alpha}^{s}$ are real \cite{LiebWu}, the ground
state is the same as that for a generalized ideal gas with
two species with above statistical interactions. In ref.
\cite{Woyn} it has been shown that solutions with complex
quasi-momenta and rapidities in the Hubbard model correspond
to gapful excitations. Therefore at low temperatures, we can
ignore thermal activations of such excitations and concentrate
on solutions with only real quasi-momenta and rapidities.

We note that our results for the Hubbard chain disagree with
those in a previous paper \cite{Spalek} on the same subject.
The disagreement comes out from the incorrect procedure taken
in Ref.~\cite{Spalek}. In the second equation of Eq.~(4) in
Ref.~\cite{Spalek}, the first and fourth terms cancel each other
exactly. Namely, the fourth term in R.H.S.,
$-\sum_{\beta=1}^M\Lambda_{\beta}\delta_{\Lambda_{\alpha},
\Lambda_{\beta}}=-\Lambda_{\alpha}$, cancels the first term.
Therefore if one proceeds correctly, the first and fourth terms on
the R.H.S. of the second equation of (5) must also cancel. For the
same reason, the first and fourth terms on the R.H.S. of (9) should
cancel as well. But surprisingly it is not the case
in Ref.~\cite{Spalek}: Two originally cancelling terms become
eventually non-cancelling!

\subsection{Strong Coupling Expansion for the Hubbard Chain}
\noindent
It is well known that spin and charge degrees of freedoms are
separated in the strong coupling Hubbard chain at zero temperature
\cite{ogata_shiba}. To explore the possibility of
a similar spin-charge separation at finite temperatures,
we first perform the $1/U$ expansion for the thermodynamic
potential of the Hubbard chain.

To begin with, we will write down the explicit form of
the thermodynamic potential $\Omega$ of the Hubbard chain.
Substituting the results of the statistical interactions
in Section~\ref{Hubbard} into the thermodynamic formulas
(\ref{Omega}) and (\ref{weq}) in Section~\ref{TFFT}, we have
the thermodynamic potential
\begin{equation}
{\Omega \over L}=-{T \over 2\pi}\int_{-\pi}^\pi
dk \log \left[1+e^{-\epsilon_c(k)/T}\right]
\label{HubOmega}
\end{equation}
for the Hubbard chain, where $\epsilon_\mu=\log w_\mu$ are
determined by the equations
\begin{equation}
\epsilon_c(k)
+{2 T\over \pi}\int_{-\infty}^\infty d\lambda^\prime
{1 \over 1+4[\lambda^\prime-(2\sin k)/U]^2}
\log \left[1+e^{-\epsilon_s(\lambda^\prime)/T}\right]
=-2\cos k -a_c ,
\label{Oepscharge}
\end{equation}
and
\begin{eqnarray}
\epsilon_s(\lambda)&-&{T \over \pi}\int_{-\infty}^\infty
d\lambda^\prime
{1 \over 1+(\lambda-\lambda^\prime)^2}
\log\left[1+e^{-\epsilon_s(\lambda^\prime)/T}\right] \nonumber \\
&+&{4 T \over \pi U}\int_{-\pi}^\pi dk
{\cos k \over 1+4[(2\sin k)/U-\lambda]^2}
\log\left[1+e^{-\epsilon_c(k)/T}\right]=0 .
\label{Oepsspin}
\end{eqnarray}

Now we proceed to the $1/U$ expansion.
When we neglect higher orders than $1/U$,
the equations (\ref{Oepscharge}) and (\ref{Oepsspin}) become
\begin{equation}
\epsilon_c(k)=-2\cos k -a_c
-{2 T\over \pi}\int_{-\infty}^\infty d\lambda
{1 \over 1+4\lambda^2}
\log\left[1+e^{-\epsilon_s(\lambda)/T}\right]
\label{epscharge}
\end{equation}
\begin{equation}
\epsilon_s(\lambda)=-{2J \over 1+4\lambda^2}+
{T \over \pi}\int_{-\infty}^\infty d\lambda^\prime
{1 \over 1+(\lambda-\lambda^\prime)^2}
\log\left[1+e^{-\epsilon_s(\lambda^\prime)/T}\right]
\label{epsspin}
\end{equation}
with the effective exchange integral
\begin{equation}
J={2 T \over \pi U}\int_{-\pi}^\pi dk \cos k
\log\left[1+e^{-\epsilon_c(k)/T}\right].
\label{exchange}
\end{equation}
The equation (\ref{epsspin}) for $\epsilon_s$ corresponds to
Eq. (\ref{eps}) of the spin-1/2 isotropic Heisenberg
chain, although $\epsilon_s$ and $\epsilon_c$ couple to each other.

Similarly we have the equations for the charge density $\rho_c$
and the spin density $\rho_s$ as
\begin{equation}
\rho_c(k)\left[1+e^{\epsilon_c(k)/T}\right]
={1 \over 2\pi}+{1 \over \pi U}\int_{-\infty}^\infty
d\lambda \rho_s(\lambda){4 \over 1+4\lambda^2}\cos k
\label{TBAeqFTcharge0}
\end{equation}
and
\begin{equation}
\rho_s(\lambda)\left[1+e^{\epsilon_s(\lambda)/T}\right]
={1 \over 2\pi} {N \over L}
 {4 \over 1+4\lambda^2}-{1 \over \pi}
\int_{-\infty}^\infty d\lambda^\prime \rho_s(\lambda^\prime)
{1 \over 1+(\lambda-\lambda^\prime)^2}
\label{spindensityHub}
\end{equation}
from (\ref{TBAeqFT}), where $N$ is the number of electrons.
Here we have neglected higher orders than $1/U$.
Comparing (\ref{spindensityHub}) with (\ref{spindensity})
of the spin-1/2 isotropic Heisenberg chain, we obtain that
the number $\rho_s$ of down spins per volume in the Hubbard chain
is proportional to the number of the electrons per volume.

Clearly, from (\ref{epscharge})-(\ref{spindensityHub}),
we conclude that the degrees of freedoms of spin and charge parts
still couple to each other at finite temperatures.
\subsection{Spin-Charge Separation in the Hubbard Chain at Low Temperatures}
\noindent
In this section, we will demonstrate that the spin-charge
separation occurs in the strong coupling Hubbard chain
at low temperatures.

Consider the case with low temperatures $T\ll 1/U$. That is
to say, temperatures are sufficiently low compared to
the effective exchange $J$ (\ref{exchange}) which behaves as
\begin{equation}
J\cong {2\over \pi U}\int_{D_-}dk \cos k [-\epsilon_c(k)]
\end{equation}
at low temperatures. Here $D_-=\{k|\epsilon_c(k)<0\}$.
Then we have $\epsilon_s=O(1/U)$ from (\ref{epsspin}).

Note that $\epsilon_c$ (\ref{epscharge}) can be rewritten as
\begin{equation}
\epsilon_c(k)=\epsilon_c^{(0)}(k)+\epsilon_c^{(1)}
\end{equation}
with
\begin{equation}
\epsilon_c^{(0)}(k)=-2\cos k -a_c
\end{equation}
and
\begin{equation}
\epsilon_c^{(1)}=-{2T \over \pi}\int_{-\infty}^\infty
{d\lambda \over 1+4\lambda^2}
\log \left[1+e^{-\epsilon_s(\lambda)/T}\right] .
\label{epscharge1}
\end{equation}
Here $\epsilon_c^{(1)}=O(1/U)$ due to the above $\epsilon_s=O(1/U)$.

{From} this observation, we devide the integral of
(\ref{HubOmega}) into two parts as
\begin{equation}
{\Omega \over L}=I_{+}+I_{-},
\end{equation}
where
\begin{equation}
I_{\pm}=-{T \over 2\pi}\int_{D_{\pm}}
dk \log \left[1+e^{-\epsilon_c(k)/T}\right],
\label{Ipm}
\end{equation}
and the ranges of integration are: $D_+
=\left\{k| \epsilon_c^{(0)}(k)+\epsilon_c^{(1)}\}>0\right\}$ and
\hfill\break
$D_-=\left\{k| \epsilon_c^{(0)}(k)+\epsilon_c^{(1)}<0\right\}$.
Since $\epsilon_c^{(0)}>>\epsilon_c^{(1)}=O(1/U)$ and $T<<1/U$,
$I_\pm$ (\ref{Ipm}) are given up to the order 1/U as
\begin{equation}
I_{+}\cong -{T \over 2\pi}\int_{D_+}
dk \log \left[1+e^{-\epsilon_c^{(0)}(k)/T}\right],
\end{equation}
and
\begin{eqnarray}
I_{-}&=&{1 \over
2\pi}\int_{D_-}dk\left\{\epsilon_c^{(0)}(k)+\epsilon_c^{(1)}\right\}
-{T \over 2\pi}\int_{D_-} dk \log
\left[1+e^{\left\{\epsilon_c^{(0)}(k)+\epsilon_c^{(1)}\right\}/T}
\right] \nonumber \\&\cong& {N \over L} \epsilon_c^{(1)}+ {1\over
2\pi} \int_{D_-}dk\epsilon_c^{(0)}(k)-{T \over2\pi}\int_{D_-} dk \log
\left[1+e^{\epsilon_c^{(0)}(k)/T}\right] \nonumber \\&=& {N \over L}
\epsilon_c^{(1)} -{T \over 2\pi}\int_{D_-} dk \log
\left[1+e^{-\epsilon_c^{(0)}(k)/T}\right].
\end{eqnarray}
where we have used the relation $\int_{D_-}dk/2\pi \cong N/ L$
which is derived from (\ref{TBAeqFTcharge0}) at low temperatures
and for large $U$. Thus we have
\begin{eqnarray}
{\Omega \over L}&\cong&
-{T \over 2\pi}\int_{-\pi}^\pi dk
\log \left[1+e^{-\epsilon_c^{(0)}(k)/T}\right]
+{N \over L}\epsilon_c^{(1)}
\nonumber \\
&=& {\Omega_c \over L}-{N \over L} {2T\over \pi}
\int_{-\infty}^\infty {d\lambda \over 1+4\lambda^2}
\log \left[1+e^{-\epsilon_s(\lambda)/T}\right]
\end{eqnarray}
for the thermodynamic potential $\Omega$ (\ref{HubOmega}) up to
the order $1/U$, where we have used (\ref{epscharge1}), and
the thermodynamic potential $\Omega_c$ of the charge part is
given by
\begin{equation}
{\Omega_c \over L}=
-{T \over 2\pi}\int_{-\pi}^\pi dk
\log \left[1+e^{-\epsilon_c^{(0)}(k)/T}\right].
\end{equation}

Thus the thermodynamic potential $\Omega$ of the Hubbard chain
can be written as the sum of the charge and the spin parts
\begin{equation}
{\Omega \over L}
\cong {\Omega_c \over L}+{N \over L} {\Omega_s \over N},
\label{scsepa}
\end{equation}
where
\begin{equation}
{\Omega_s \over N}= -{T\over 2\pi}
\int_{-\infty}^\infty d\lambda {4 \over 1+4\lambda^2}
\log \left[1+e^{-\epsilon_s(\lambda)/T}\right],
\end{equation}
which is identical to the thermodynamic potential (\ref{OmegaXXX})
of the spin-1/2 isotropic Heisenberg chain. Eq. (\ref{scsepa}) is an
indication of charge-spin separation.

\section{Exactly Solvable Model in Higher Dimensions}
\label{HKmodel}
\noindent
So far we have discussed the models in one dimension to which
the Bethe ansatz approach is applicable. In two dimensions,
it is known that the quasiparticles of the fractional quantum Hall
liquid are anyons. They can also be considered to be exclusons.\cite{Haldane}
In this section we present an example of mutual
exclusion between different species in an exactly solvable model
in {\em higher dimensions}, that exhibits charge-spin separation
under certain circumstances. This clearly shows that exclusion
statistics is conceptually different from anyon statistics
whose existence requires two (spatial) dimensions.

Recently two of us \cite{HK} and Baskaran \cite{Baskaran} have
proposed a model of interacting electrons that can be solved
exactly in any dimensions. The Hamiltonian is
\begin{equation}
H= -\sum_{\langle i,j\rangle}c_{i,\sigma}^\dagger c_{j,\sigma} + h. c.
+{\frac U {L^d}}\sum_{i,j,\ell,m}\delta_{i+j,\ell+m}
c_{i,\uparrow}^\dagger c_{j,\uparrow}
c_{\ell,\downarrow}^\dagger c_{m,\downarrow}
- \sum_{i,\sigma} (\mu + \sigma \mu_0 h )
c_{i,\sigma}^\dagger c_{i,\sigma},
\label{HKHamil}
\end{equation}
where $\langle i,j\rangle$ represents nearest neighbors in $d$
dimensions, and $L^d$ is the total number of lattice sites,
$\mu$ the chemical potential, $\mu_0$ the magnetic moment and $h$
the external magnetic field.

This model is unrealistic in the sense that the interaction term
with coefficient $U$ is of infinite-ranged in real space and of
strength independent of distance. (Note that the noninteracting electon
models are even more unrealistic since they neglect the long-range
Coulomb.) It, however, has an attractive
feature of being exactly solvable. In fact, it can be easily
diagonalized for each $k$ in momentum space. All the properties
including the thermodynamic quantities were obtained
in \cite{HK}. Furthermore
it is remarkable that this model exhibits a number of important
features of the correlated electron problems in spite of
its simplicity and unrealistic nature. It exhibits, for example,
the Mott metal-insulator transition that was stressed by two of
us \cite{HK} and  Continentino and Coutinho-Filho \cite{Brazilpaper}.

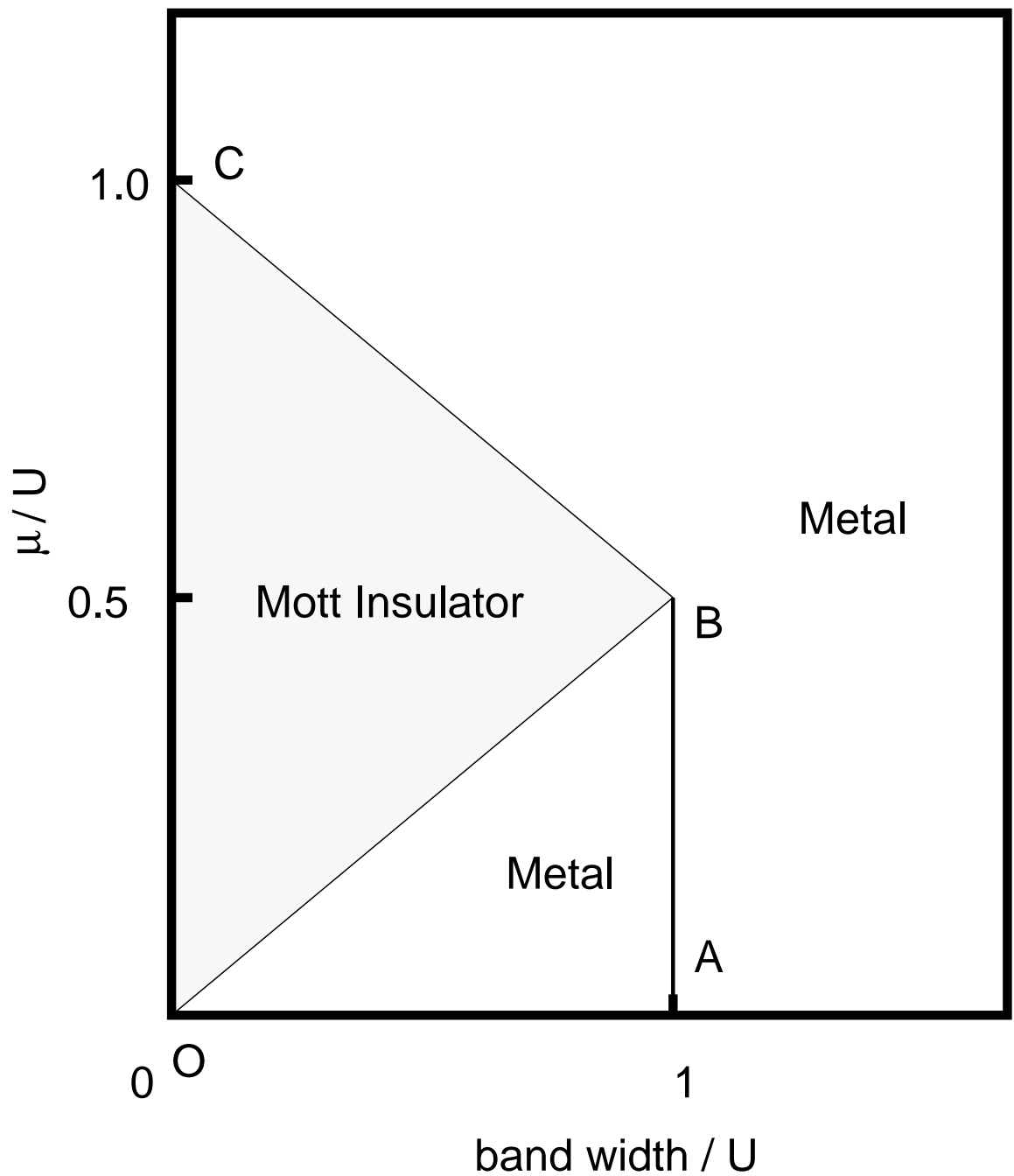
\begin{figure}
\caption{
Phase diagram of the exactly solvable model in higher dimensions at $T=0$.
 \label{fig}}
\end{figure}

The zero temperature phase diagram of this model in any dimensions
is shown in Fig.~\ref{fig}. It has both the fixed-density and
density-driven Mott transitions. These transition in general may
be in different universal classes \cite{Continentino,Fisheretal}.
The critical exponents of the two types of transitions are, however,
the same in the present model and they seems to be in the same
universality class \cite{Brazilpaper}. In the zero temperature
phase diagram Fig.1, the region OBC is a Mott insulator phase with
half filled band. The rest is metallic phases. In the region OABC,
a double occupancy is prohibited and, as we will show, the
system can be described by an excluson picture.
There is a Fermi surface of the excluson gas in the region OAB ,
and on the phase transition line OB is a quantum phase transition
of the excluson gas. On the other hand, the excluson description
breaks down on the phase transition line BC. The region outside
OABC is a metallic phase which is described by the two species of
fermions (spin up and spin down electrons) as it should be.

Let us concentrate on the region OAB. Assume that $U$ is large
and $T$ is low, so that $U$ is much larger than both $T$ and
the band width \cite{SpalekWojcik}. Under these conditions,
there is no activation
of doubly occupied states, therefore there are only
three states (0,1 and 2) for each momentum $k$.
In the state 0 there is no electron, the state 1 an electron
with spin up, and the state 2 an electron with spin down.
Let us denote the number of charges as $N_c$ and the number
of magnons (number of spin-down) as $N_s$. We regard $N_c$
and $N_s$ as independent variables (spin-charge separation).
By definition, the state 0 has $N_c=0$ and $N_s=0$, the state 1
$N_c=1$ and $N_s=0$ and the state 2 $N_c=1$ and $N_s=1$
(See table~\ref{table1}). Then from (\ref{mutualEq}), we
easily derive
\begin{equation}
G_c=1 \ , \ G_s=0,
\end{equation}
and
\begin{equation}
\left[
\matrix{g_{cc}&g_{cs}\cr
g_{sc}&g_{ss}}
\right] =
\left[
\matrix{1&0\cr-1&1}
\right] \; .
\label{gHK}
\end{equation}

It is easy to verify that the condition for the ideal excluson
gas (\ref{totalE}) is satisfied, so that the system with doubly
occupied states suppressed can be described as a generalized ideal
gas with two (charge and magnon) species with the statistical
interaction given by (\ref{gHK}). Note the nontrivial value $-1$
for mutual statistics $g_{sc}$; i.e.\  the presence of a charge
can creat a  magnon state, though there is no bare
available single magnon state ($G_{s}=0$) when there is no charge.
It is straightforward to check that the thermodynamics of the
generalized ideal excluson gas obtained from Eqs.~(\ref{eq_det_w})
and (\ref{OmegaExdes})
is identical to the result of Ref.~\cite{HK} in the low
temperature limit.

Indeed, in the present case, the species index $\mu=c,s$, and
the state index $j$ is the momentum $k$ in $d$-dimensional space.
Equation~(\ref{OmegaExdes}) now takes the form
\begin{equation}
\Omega = -T\sum_\mu\int
{\frac {d^dk}{(2\pi)^d}}
\log\left[
\frac
{1-n_\mu(k)-\sum_\nu g_{\mu\nu} n_\nu(k)}
{1-\sum_\nu g_{\mu\nu} n_\nu(k)}\right],
\label{OmegaHK}
\end{equation}
where $n_\mu(k)$ is the occupation number distribution function of
the charge ($\mu=c$) or spin ($\mu=s$) excitations in
$k$-space. From Eqs.~(\ref{eq_for_w}) and (\ref{eq_det_w}), we have
\begin{equation}
n_c(k)[1+w_c(k)] = 1, \ \ n_s(k)[1+w_s(k)] = n_c(k),
\label{HKn}
\end{equation}
where the statistics matrix (\ref{gHK}) has been used , and
$w_c(k)$, $w_s(k)$ satisfy
\begin{eqnarray}
w_c(k) {\frac{1+w_s(k)}{w_s(k)}} &=& e^{(\epsilon_c(k)-\mu_c)/T}, \\
w_s(k) &=& e^{(\epsilon_s(k)-\mu_s)/T},
\label{HKws}
\end{eqnarray}
where $\epsilon_c(k)=-2\sum_{\alpha=1}^d \cos(k_\alpha) $,
$\epsilon_s=2\mu_0h$
(the energy  of spin excitation, which is actually
measured relative to the energy of spin-up electrons), and
\begin{equation}
\mu_c=\mu+\mu_0 h,\;\; \mu_s=0.
\end{equation}
Thus
\begin{equation}
w_c(k)=\frac
 { e^{(\epsilon_c(k)-\mu_c)/T}} {1+e^{-(\epsilon_s(k)-\mu_s)/T}}.
\label{HKwc}
\end{equation}
Substituting Eqs.~(\ref{HKwc}), (\ref{HKws}), and (\ref{HKn}) into
(\ref{OmegaHK}),
we obtain
\begin{eqnarray}
\Omega&=&-T\int {\frac {d^dk}{(2\pi)^d}}
\log\left[1+(e^{\mu_0h/T} + e^{-\mu_0h/T})
e^{(\mu_c-\epsilon_c)/T}\right] \nonumber \\
&=&-T\int {\frac {d^dk}{(2\pi)^d}}
\log\left[1+e^{-(\epsilon_1(k)-\mu)/T}
+ e^{-(\epsilon_2(k)-\mu)/T}\right],
\label{HKresult}
\end{eqnarray}
where $\epsilon_1(k)=\epsilon_{c}(k)-\mu_0h$,
and $\epsilon_2(k)=\epsilon_{c}(k)+\mu_0h$ are
the energy of spin-up and spin-down electrons respectively.
Equation~(\ref{HKresult}) is nothing but the result of
Ref.~\cite{HK} in the low temperature limit.
(Remember that here we consider
the case with large $U$ and low $T$, so that doubly occupied states
are suppressed.)

Here we emphasize that the concept of spin-charge separation
is crucial. The effects that spin and charge excitations
are not actually independent of each other have been taken care of
by the statistical interaction or mutual statistics between them
in the present formulation. It seems to us that similar situations
may happen in other strongly correlated systems that exhibit
charge-spin separation in higher dimensions. We finally note that
our treatment for the exactly solvable models in higher dimensions
disagrees with that in a previous paper \cite{Spalek}
on the same subject.

\section{Acknowledgements}
This work was supported by Grant-in-Aid from the Ministry of
Education, Science and Culture of Japan.
The work of Y.S. Wu was also supported in part by U.S. NSF grant
PHY-9309458. He is also grateful for  warm
hospitality of the Institute for Solid State Physics, University of
Tokyo during his visits.


\begin{table}
\caption{Electronic states and spin charge labels
\label{table1}}
\begin{tabular}{cccc}
Electronic State & Label & {$N_c$: Charge } & {$N_s$: Magnon} \\
\tableline
$0$&$0$&$0$&$0$\\
$\uparrow$&$1$&$1$&$0$\\
$\downarrow$&$2$&$1$&$1$\\
\end{tabular}
\end{table}

\end{document}